\documentclass[aps,prl,showpacs,preprintnumbers,onecolumn,twocolumn,nofootinbib,superscriptaddress]{revtex4-1}

\usepackage{geometry}                % See geometry.pdf to learn the layout options. There are lots.
\usepackage{graphicx}
\usepackage{amssymb}
\usepackage{dsfont}
\usepackage{epstopdf}
\usepackage{color}
\usepackage{mathtools}
\usepackage{nicefrac}
\usepackage{float}
\usepackage[caption = false]{subfig}
\usepackage{hyperref}
\usepackage{cleveref}
\definecolor{red}{rgb}{1,0,0}
\definecolor{blue}{rgb}{0,0,1}
\definecolor{black}{rgb}{0,0,0}

\usepackage[dvipsnames]{xcolor}

\newcommand{\p}{\partial}
\newcommand{\eq}[1]{\begin{align}#1\end{align}}
\newcommand{\eqs}[1]{\begin{align*}#1\end{align*}}
\newcommand{\ffrac}[2]{\mbox{$\frac{#1}{#2}$}}
\newcommand{\half}{\mbox{$\frac{1}{2}$}}

\newcommand{\TT}{\mathcal{T}}
\newcommand{\FF}{\mathcal{F}}
\newcommand{\GG}{\mathcal{G}}
\newcommand{\ZZ}{\mathbb{Z}}

\usepackage{calrsfs}
\usepackage{comment}
\usepackage{scalerel,stackengine}
\stackMath
\newcommand\widecheck[1]{%
\savestack{\tmpbox}{\stretchto{%
  \scaleto{%
    \scalerel*[\widthof{\ensuremath{#1}}]{\kern-.6pt\bigwedge\kern-.6pt}%
    {\rule[-\textheight/2]{1ex}{\textheight}}%WIDTH-LIMITED BIG WEDGE
  }{\textheight}% 
}{0.5ex}}%
\stackon[1pt]{#1}{\scalebox{-1}{\tmpbox}}%
}

\newcommand{\PP}{\mathbb{P}}

\newcommand{\ed}{\tilde\epsilon_d}
\newcommand{\es}{\tilde\epsilon_s}

\makeatletter
\renewcommand\paragraph{\@startsection{paragraph}{4}{\z@}%
    {1.5ex \@plus1ex \@minus.2ex}%
    {-1em}%
    {\normalfont\normalsize\bfseries}}
\makeatother

\begin{document}
\title{Phase structure of the Random Language Model}
\author{Alessio Giorlandino}
\affiliation{Department of Physics, International School of Advanced Studies, Trieste, Italy}
\author{Eric De Giuli}
\affiliation{Department of Physics, Toronto Metropolitan University, M5B 2K3, Toronto, Canada}
\author{Sebastian Goldt}
\affiliation{Department of Physics, International School of Advanced Studies, Trieste, Italy}
%\date{\today}                     

\begin{abstract}
  Context-free grammars are minimal models of hierarchical structure in
  human language, generating structured text from recursive production rules.
  The Random Language Model (RLM)~\citep{DeGiuli19}, an ensemble of such
  grammars with random rule weights, exhibits a cross-over from
  gibberish-like output to structured text as a function of a ``temperature'',
  but the location and nature of this transition remained
  unclear. Here, we show that the RLM exhibits a hierarchy of phase
  transitions in a double-scaling limit where the grammar temperature $\ed \to
  0$ and the number of hidden symbols $N \to \infty$ at fixed $x = \ed \log N$.
  By identifying the relation between RLM and the Random Energy Model, we
  identify a series of transitions where correlations between visible symbols
  emerge, single-symbol marginals become nonuniform, and rule use freezes in a
  glassy phase. A semi-annealed approximation yields nontrivial scaling laws for
  rule usage, entropy, and energy, consistent with Heaps' law and context-length
  scaling observed in large language models.
\end{abstract}

\maketitle

The ubiquity of Large Language Models (LLMs) has made the
quantitative understanding of language an urgent problem. Is mastery of language
sufficient for artificial general intelligence, or a subtle mirage? Do scaling
laws of LLM performance \cite{Kaplan20}
reflect universal properties of language?
Can first language acquisition in
humans be understood within the same framework as training of artificial neural
networks? To address such questions, one of us \cite{DeGiuli19} introduced the Random
Language Model (RLM), an ensemble of context-free grammars, which captures the universal hierarchical
property of human and computer language \cite{DeGiuli19, DeGiuli19a, De-Giuli22,
Nakaishi22, Lalegani24}. Related models of language have since been used to
study the capabilities and limitations of neural networks~\cite{Lin17,
Cagnetta24, Cagnetta24a, Garnier-Brun24, cagnetta2026deriving}. 

\begin{figure}[b!]
  \includegraphics[width=.8\columnwidth]{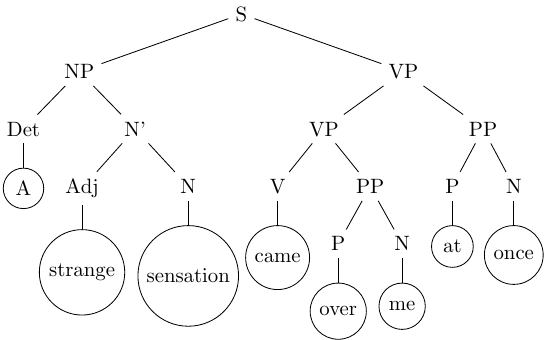}
  \caption{\label{fig_tree} Example derivation for an English sentence in a
    context-free grammar. Terminal symbols are circled; hidden symbols denote
    parts of sentences or specific word types. }
\end{figure}

Briefly, a context-free grammar (CFG) is a system of rules for producing
structured strings from a finite alphabet~\cite{Chomsky02,Chomsky14}. It
consists of a set of hidden, or non-terminal, symbols; a set of observable, or
terminal, symbols; and production rules that replace one non-terminal symbol by
either a pair of non-terminals or by a terminal symbol. Without loss of
generality, we work in the Chomsky normal form~\citep{Hopcroft07}, where all
rules are of the form $a \to b c$ or $a \to A$, where $a, b, c$ are hidden
symbols and $A$ is a visible, terminal symbol. Starting from the root symbol, repeated application of these rules generates a derivation tree whose leaves form an observable sentence like the one shown in \cref{fig_tree}. The study of context-free grammars has been a cornerstone of linguistics and computer science and has found
applications in biochemistry, compiler design, and quasicrystals, among others.

The key idea of Ref. \cite{DeGiuli19} was to study a statistical \emph{ensemble} of
grammars. Each production rule is assigned a positive weight: bulk weights
$X_{abc}$ for rules $a\to bc$ inside the derivation tree, and boundary weights
$Y_{aA}$ for rules~$a\to A$ which control the emission of observable symbols.
For a fixed derivation tree like the one shown in \cref{fig_tree}, the weight of
a complete derivation is the product of the weights of all the rules
appearing in the tree, defining a Gibbs measure over hidden and observed
configurations. Drawing weights from a lognormal distribution, Ref. 
\cite{DeGiuli19} demonstrated that the resulting grammars go from a
``babbling'' phase, where the generated text is unstructured and its entropy is
maximal, to a structured phase where the entropy of the text drops
precipitously. These phases are spanned by a temperature-like parameter
$\tilde\epsilon_d$ controlling the distribution of the grammar rules.

While Refs.~\cite{DeGiuli19,DeGiuli19a,De-Giuli22} showed that this transition is
captured by a spin-glass order parameter adapted to trees, the question and
possible location of a true thermodynamic phase transition was left open.
Numerical results~\cite{DeGiuli19} showed a scaling collapse over the number
of hidden symbols, $N$, but with a modest logarithmic dependence on $N$.
Observables did not distinguish a single critical temperature.

\begin{figure*}[ht!]
  \includegraphics[width=\linewidth]{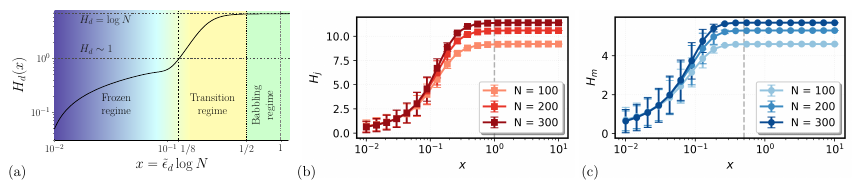}
  \caption{\label{fig:phase-diagram} (a) Phase diagram of the RLM as a function
  of rescaled temperature $x=\ed\log N$. The schematic behavior of the entropy
  rate $H_d$ is indicated. (b,c)
  Destabilization of the `paramagnetic' solution: considering a single site, (b) the entropy of the joint distribution $H_j$ of two symbols begins to drop at $x
  \approx 1$ (dashed) while (c)
  the entropy of the marginal distribution $H_m$ of a single symbol begins to drop at $x \approx 1/2$ (dashed).}
\end{figure*}

Here, we show that the RLM exhibits a surprisingly rich phase structure that
obscured previous results. We exploit the limit $N \to \infty$, which linguistically corresponds to adding features to symbols, such as case, tense, gender, and so on. By considering a scaling limit in which $\ed \to 0$
while $N \to \infty$ at finite $x = \ed \log N$, we show that the RLM undergoes
a series of transitions. Starting from high temperatures, where all symbols are equally likely, we first identify the temperatures at which correlations between symbols emerge and marginals become non-uniform,
respectively. We identify these transitions by analysing the belief-propagation (BP)
equations for a given derivation tree, which reveals a connection between the RLM
and the Random Energy Model of \citet{Derrida80,Derrida81}. To analyze the behavior at even lower temperatures, we develop a
``semi-annealed'' approximation, in which imperfect rule sampling at finite $N$ and the heavy-tailed nature of rule distributions at small $\ed$ are both respected. The semi-annealed theory describes the phase transition at $x_c=1/8$ where the entropy rate $H_d$ of the generated text drops, which corresponds to the phase transition from babbling to structured text. Our analysis is summarized in the phase diagram of
\cref{fig:phase-diagram}a.
The semi-annealed theory also explains scaling laws in the statistics of rule use.
We finally make the connection to natural language by recovering Heaps
law~\cite{Heaps78} of vocabulary growth, and by rationalising the dependence of
entropy on context length in LLMs~\cite{Scheibner25}. An extended version of the theory is presented in Ref.~\cite{De-Giuli26}.

\paragraph{Model definition} The RLM defines the energy of a
derivation with hidden symbols $\sigma$ and visible symbols $o$ on a tree $\TT$ as \footnote{Here $E_0$ is an unimportant constant. We have $E_0 = (\ell-1) \log
N^2 + \ell \log T$, where $\ell$ is the number of leaves of $\TT$ and $T$ is the
number of observable symbols in the grammar.}
\begin{equation}
  \label{eq:rlm-energy}
  E = E_0-\sum_{a,b,c} \pi_{abc} (\sigma; \TT) X_{abc} -\sum_{a,B} \rho_{aB} (\sigma,o; \TT) Y_{aB}
\end{equation}
where $X_{abc} (Y_{aB})$ is the log-weight of rule $a\to bc$ ($a \to B$) and
$\pi_{abc}(\sigma; \TT)$ counts the number of times rule $a \to bc$ is used in
the derivation $\sigma$, and likewise for $\rho_{aB}$. Trees are sampled by a branching process with emission probability $p_E$, whose value is chosen to fix the typical tree size.  
The RLM defines an ensemble over grammars in which rule weights $\{ X_{abc} \}$ and $\{ Y_{aB} \}$ are centered Gaussian random variables, with variances $\nicefrac{1}{2\ed}$ and $\nicefrac{1}{2\es}$, respectively. Large $\ed$ corresponds to small rule-weight variance, and hence a large variability in outputs, since many rules have similar weights. In this respect, $\ed$ behaves like a temperature, which motivates our definition of rescaled temperature $x = \ed \log N$; see \cite{Lalegani24} for a formal mapping to temperature. 

\paragraph{Emergence of local correlations}
To detect the emergence of structure in the text generated by a random
context-free grammar at a given temperature, we analyse the belief propagation
equations (BP) \citep{Mezard09} for the distributions of hidden symbols
on the tree. At high temperatures, all
messages are uniform, corresponding to a structure-less, paramagnetic state in
which each symbol is equally likely.
As we decrease the temperature, the paramagnetic state loses its
stability under the BP update and structure emerges. Analysing the stability of
BP fixed points is a powerful tool for uncovering thresholds at which structure
in a random object appears~\citep{mezard2002analytic,
braunstein2005survey, braunstein2006learning,
zdeborova2007phase, decelle2011inference, krzakala2012statistical, zdeborova2016statistical}. We perform this analysis at a
generic internal factor $f$, applying a rule $a \to bc$ with parent $a$ and
children $b$ (left) and $c$ (right). 

The weakest structure such a factor can generate is a correlation between its
two children while each child's marginal remains uniform. By holding
the message reaching $f$ from above uniform, we can examine the joint distribution
it induces on the pair $(b,c)$ %. Unlike a standard BP message, which is a
% function of a single variable, this is a pair-level object that captures the
% correlation $f$ induces between the two siblings 
through their shared dependence
on $X$. Summing out the parent, we have
\begin{equation}
  \label{joint}
  \psi_{f \to (\text{left},\text{right})}(b, c) = \frac{\sum_{a=1}^{N} e^{ X_{abc}}}{\sum_{a,b',c'=1}^{N} e^{ X_{ab'c'}}},
\end{equation}
which we recognise as a ratio of two Random Energy Model-like partition
functions~\cite{Derrida80}, with $N$ terms in the numerator (only the parent
index is summed out) and~$N^3$ terms in the denominator.
The limiting behavior of sums of exponential random variables was rigorously
established by \citet{Ben-Arous05} and undergoes two transitions governed by the number $k$ of independent
indices summed over (the sum runs over $N^k$ terms): the law
of large numbers holds for $x > x_{\mathrm{LLN}}(k) = \nicefrac{1}{4k}$, and within that
range the Central Limit Theorem (CLT) holds in the narrower window $x >
x_{\mathrm{CLT}}(k) = \nicefrac{1}{k}$. The numerator, with $k=1$, is the bottleneck: it
satisfies a CLT only for $x > x_{\mathrm{CLT}}(1) = 1$. In this regime, the joint
message is uniform, $\nicefrac{1}{N^2}$, with sub-leading Gaussian fluctuations. For $x < 1$
the CLT breaks down in the numerator: fluctuations become heavy-tailed and the
joint message deviates from $\nicefrac{1}{N^2}$, signalling that $f$ broadcasts genuine
correlations between the two children. This first transition marks the emergence
of \emph{local structure}: siblings sharing a common parent begin to be
correlated with one another, even though the marginal distribution of each node
remains uniform. We confirm this numerically in \cref{fig:phase-diagram}(b): for
$x > 1$, the entropy of the joint $H_j = 2 \log N$, its maximum value for a uniform
distribution over symbols, and begins to decrease at $x_{\mathrm{local}} =
1$. However, the marginal distribution over symbols remain uniform -- they
acquire structure in a second transition, as we show next. 
%Strong finite-size effects smooth the transition, so at accessible sizes it appears as a crossover rather than a sharp phase transition.
%\EdG{Referees will jump on this. Fig 4b shows that with the right observable, a phase transition is evident at modest N. I think nothing is gained by saying that 1/2 and 1 are phase transitions, since it invites a request for proof, via change in an order parameter, etc. We need to use a more careful language}

\begin{comment}
\begin{equation}
  \label{bp-general}
  \psi_{f \to \text{left}}(a) = \frac{\sum_{r,c=1}^{N} e^{\beta X_{rac}}\, \chi_{\text{root} \to f}(r)\, \chi_{\text{right} \to f}(c)}{\sum_{\ell,r,c=1}^{N} e^{\beta X_{r\ell c}}\, \chi_{\text{root} \to f}(r)\, \chi_{\text{right} \to f}(c)},
\end{equation}
\end{comment}

\paragraph{Emergence of non-uniform marginals} To analyse the marginal
distribution of a symbol in a given node, we analyse a single downward BP
message. Focusing again on a generic internal factor $f$ applying a rule $a \to bc$ with parent
$a$ and children $b$ (left) and $c$ (right), the message that $f$
sends to the left child
depends on the incoming messages from the parent and the right child. These
incoming messages depend on the rest of the tree, but since we are testing the 
stability of the \emph{uniform} configuration, we can set both incoming messages to $1/N$ and ask whether the outgoing message stays uniform in turn. Under this ansatz, the message reads
\begin{equation}
  \label{marg}
  \psi_{f \to \text{left}}(b) = \frac{\sum_{a,c=1}^{N} e^{ X_{abc}}}{\sum_{a,b',c=1}^{N} e^{ X_{ab'c}}}.
\end{equation}
This is the message the left node propagates onward to the rest of the tree
(and, under the uniform ansatz, its own marginal). % The question of strong
% structure is precisely when \eqref{marg} stops being uniform.
The message is again a ratio of REM-like partition functions, with $N^2$ terms
in the numerator and~$N^3$ in the denominator. The CLT now governs the numerator
down to $x_{\text{CLT}}(2) = 1/2$, a lower threshold than the joint case: the
extra summed index means more independent terms, so concentration survives to
smaller $x$. For $x > 1/2$ both sums concentrate and \eqref{marg} equals~$\nicefrac{1}{N}$
up to vanishing Gaussian fluctuations: the message is uniform and so are the
marginals. For $x < 1/2$ the CLT breaks down in the
numerator, \eqref{marg} deviates from $\nicefrac{1}{N}$, and structured messages
propagate down the tree yielding non-uniform marginals. We confirm our prediction numerically by computing the
marginal entropy $H_m$ of the left child in a single triple ---
equivalent to the full tree under the uniform-message assumption --- which departs from its maximal value around $x = 1/2$, see
\cref{fig:phase-diagram}(c).

This second threshold marks the onset of a \emph{transition} regime where the
entropy rate continues to decrease with temperature until a phase transition
occurs at $x = 1/8$. At that temperature, the tree becomes too constrained to sustain a diverging
entropy rate: it drops from $O(\log N)$ to $O(1)$, signaling the onset of the
frozen phase, which we study next. These transitions are summarized in
\cref{fig:phase-diagram}(a), along with a schematic of the entropy rate $H_d$,
to be elucidated below. 

\begin{figure*}[t!]
\includegraphics[width=\textwidth]{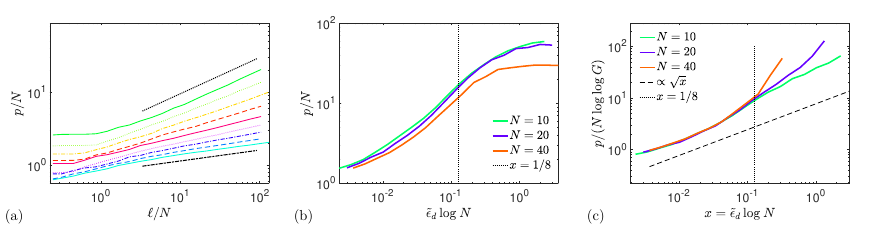}
\caption{ Rule use as a function of (a) corpus length and rescaled temperature (b,c). (a) Number of unique rules $p$ per symbol in a corpus of length $\ell$, at $N=40$ and $\ed$ from $10^{-3}$ (blue, bottom) to $10^{-1.6}$ (green, top), equivalent to $x = 4 \times 10^{-3}$ to $x=0.093$.  Heavy dotted and dash-dotted lines indicate powers $\ell^{0.5}$ and $\ell^{0.15}$, respectively. (b) $p/N$ versus $x=\ed\log N$ at indicated $N$, at fixed $n\approx 850$. (c) In the frozen regime, the remaining $N$ dependence is approximately collapsed by a factor of $\log \log G$. The vertical dotted line indicates the freezing transition $x=1/8$. The black dashed line shows $p \propto \sqrt{x}$, predicted for $x \ll 1/8$.
}\label{fig3}
\end{figure*}

\paragraph{Patterns and multiplicity} To build a quantitative theory of the RLM
in the frozen phase, we aim to identify relevant classes of excitations. Similar
to how spin-wave and vortex excitations make manifest the criticality of the XY
model \cite{Kosterlitz73, Kosterlitz74}, we first rewrite the partition function
in terms of excitations called patterns. Throughout, we assume that the surface
layer is deterministic; this is equivalent to assuming that each hidden symbol
has a unique terminal.

Fix a set of $m$ trees, called a forest~$\FF$ and
consider the partition function of the RLM on $\FF$, $\ZZ(\FF; \GG) = \sum_{\{
\sigma, o \} } e^{-E(\sigma,o)}$, where $\GG = (X,Y)$ defines the grammar, and
$\sigma (o)$ are the hidden (observable) symbols, respectively. The sum over forests is
controllable and benign; see \cite{DeGiuli19,DeGiuli19a,De-Giuli22}. If a given rule
$a \to bc$ appears $\pi$ times, then its weight appears in the sum as $e^{\pi
X_{abc}}$. Since the breakdown of the LLN and CLT depend on the variance of rule
weights, $\ZZ$ will be very sensitive to the multiplicity of rules. Define the
rule spectrum $\{p_k, k \geq 1\}$ that for each $k$ indicates how many rules are
used $k$ times, with~$p_k$ counting the number of hidden rules. For
example, the tree in Fig.~\ref{fig_tree} has 7 branches, of which 2
use the same rule, so that $p_1 = 5, p_2 = 1,$ and $p_k=0$ for~$k \geq 3$. The
set of all derivations consistent with a given spectrum is called a
\textbf{pattern}, $P(\{p_k\}; \FF)$. These are the excitations of the model.
Patterns coarse-grain over microscopic derivations by retaining only rule usage
frequencies, analogous to occupation numbers in statistical mechanics. 

In a forest of $m$ trees of total length $\ell$, the number of branches 
$n=\ell-m$ is fixed by topology, so the most elementary observable is $p$, the total number of unique hidden rules used in the pattern. It plays a central role, as it encodes the text structure as a function of corpus size $\ell$, grammar size $N$, and reduced temperature $x$. 

%In a forest of $m$ trees of total length $\ell$, there are $n=\ell-m$ branches and $n = \sum_{k \geq 1} k p_k$. The number of unique rules $p = \sum_j p_j$ emerges as a key observable in the theory. 

Empirically, we find
that $p$ displays nontrivial behaviors as functions of the number of branches $n$, the rescaled temperature
$x$, and the number of hidden symbols $N$, as shown in \cref{fig3}. In the frozen phase, we find $p \sim n^{\beta(x)}$ where the exponent $\beta(x)$ continuously varies with $x$, with $\beta(1/8) \approx 0.5$ and $\beta(0.004) \approx 0.15$. At fixed $n$ and small~$x$, we have $p \sim x^{0.5}$; 
this steepens as $x \to 1/8^-$. Finally, at fixed $n$ we find that the $N$
dependence is captured by $p \propto N \log \log G(n,x,N)$ where $G = (n/Na)$ is
a rescaled corpus length that emerges from the theory. Here $a= \sqrt{2\ed w} \sim \sqrt{8x}$ and $w = W_0(N^4/2\pi)$, where $W_0$ is Lambert's $W_0$. The semi-annealed
approximation that we introduce next can rationalise all of these findings.
 
\paragraph{The semi-annealed approximation} The partition sum $Z$ can be organized as a sum over patterns, viz., 
\eq{ \label{Z}
%Z(\TT; \GG) = \sum_{\{ P(\TT) \}} \underbrace{\sum_{ (\sigma,o) \in P } e^{\beta E(\sigma,o; \GG)}}_{\mathcal{P}(P; \GG)} %Z_P(\sigma,o,\GG)
\ZZ(\FF; \GG) = \ZZ_0 \sum_{\{ P  \}} \underbrace{\sum_{ (\sigma,o) \in P } e^{- E'(\sigma,o; \GG)}}_{\mathcal{P}(P; \GG)} %Z_P(\sigma,o,\GG)
}
%where $\ZZ_0 = Z_{tree}^{-m} (p_E \overline{O})^{\sum_i \ell_i} ((1-p_E)\overline{M})^{-m+\sum_i \ell_i}$ 
where $E'=E-E_0$ and $\ZZ_0 = \prod_i Z(\ell_i)$ depends only on the number of trees $m$ and the total length $\ell = \sum_i \ell_i$. 
%\textcolor{gray}{For any derivation, applying a global permutation to the symbols results in another derivation with the same pattern, but different rules. Since the model is defined by iid rule weights, each pattern thus explores the space of grammars. %In other words, we can evaluate the contribution of each pattern In other words, even though $\ZZ$ is computed for a fixed grammar, sums over symbol values induce sums over possible grammar weights. It is however essential to respect the rule spectrum of the pattern, since this controls the Boltzmann weight.}
 Each pattern gains a huge amount of entropy over individual derivations, so that $\ZZ$ has a large-deviation structure \cite{De-Giuli26}.  %Hence they can be estimated by appropriate annealed 
Since patterns are invariant under permutations of rule indices, the dominant saddle-point depends only on the statistics of rule weights and not on their specific realization, justifying an average over disorder at fixed multiplicity structure. 
At finite $N$, this average must be restricted by extreme-value statistics, since only a finite portion of the weight distribution is sampled. 
This motivates a {\it semi-annealed approximation,} in which we estimate $\ZZ$ by a restricted `annealed' average over grammar weights. % but also taking into account that the true $\ZZ$ is computed at a fixed $\GG$ and finite $N$.
%, which means that it does not explore its entire probability space. 
We expect that the approximation, which considers the most likely pattern, captures the leading behavior of $\ZZ$ in the scaling limit.

There are two main ingredients. First, since trees can be sampled sequentially from the root without backtracking, for a given head $a$ there are $N^2$ possible rules $a \to bc$ to consider. This defines a REM partition sum for each branch.
%First, we must account for the correlations that are induced by matching symbols on the tree: the daughter of one branch is the head of the next branch, and so on. Since we can sample trees in a sequential fashion, from the root downwards, this amounts to computing a partition function for a branch, conditional on its head value: this reduces the number of DOF down to $N^2$ per branch.
Second, although rule distributions are heavy-tailed at small $\ed$, at any finite $N$ we only sample a finite number of values from the distribution. This defines a value $X_*$ such that the rule distribution is only typically sampled for $X<X_*$. 

Then in the semi-annealed approximation, a pattern contributes
\eq{ \label{P}
\mathcal{P}(\{p_k\}) = \Omega N^Q \prod_k \langle e^{k X} \rangle_{X<X_*}^{p_k} 
}
where $Q$ counts the number of DOF in the average over symbols, $\Omega$ accounts for the entropy of the pattern, and~$\langle \cdot \rangle_{X<X_*}$ is a restricted rule average over $X<X_*$. In the regime $p>N$ relevant for a large corpus, $\Omega$ can be estimated by counting degreees of freedom; an exact computation with the RLM field theory \cite{DeGiuli19a,De-Giuli22} confirms and extends this result \cite{De-Giuli26}. The derivation, presented in End Matter, obtains an expression for $\Omega$ and $Q=3p$. Moreover, it identifies a correlation length $\ell_c$ such that the total text with $n$ branches has $m_c=n/\ell_c$ correlation volumes. The latter is $m_c=p/N$ in terms of the number of unique rules.

Within the frozen regime, the solution distinguishes further regimes: a critical regime close to $x=1/8$; a saturation regime for $x \lesssim x_{\mathrm{lo}}$, with $x_{\mathrm{lo}} \sim 10^{-7} (n/N)^{-10/3}$, and a scaling regime in between. The latter is so-called because $n$ and $N$ nontrivially affect observables only through $G=n/(Na)$ where $a \sim \sqrt{8x}$.

\begin{figure*}[t]
\includegraphics[width=\textwidth]{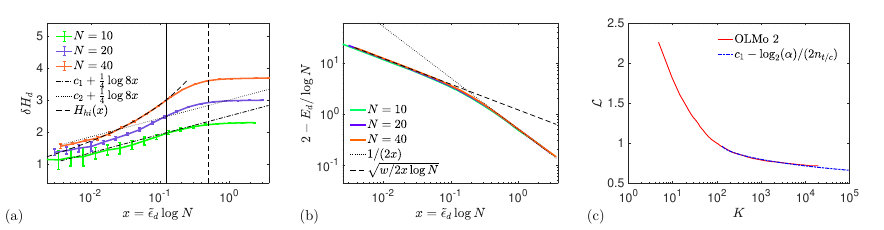}
\caption{ Entropy, energy, and code length. (a) Entropy rate in the RLM at indicated $N$ along with theoretical predictions: logarithmic law deep in the frozen phase (dash-dotted, dotted lines) along with critical corrections $H_{hi}(x)$ for $x<1/8$ (dashed). Vertical lines indicate regime boundaries $x=1/8$ and $x=1/2$. (b) Rescaled energy at indicated $N$ along with theoretical predictions in the frozen (dashed) and transition (dotted) regimes. Here $w=W_0(N^4/2\pi)\sim 4\log N$. (c) Code length in OLMo 2 as a function of context length (red) \cite{Scheibner25}, and theoretical prediction (blue) from the RLM. The blue curve has the leading behavior $\sim \text{const}-\half \log \log K$ in nats/symbol (see text). 
}\label{fig4}
\vspace*{-1em}
\end{figure*}

After identifying the dominant pattern (see \cite{De-Giuli26}), we find that $p$ has the leading behavior in the frozen phase
\eq{ \label{p}
%p/N & \approx \frac{2\sqrt{2\pi} e^{3/2}}{3} a   \log(\alpha) G^{\frac{2}{1+\alpha}}
p/N & \approx a \sqrt{4\pi e^3/ 9}  \log(\alpha) G^{\frac{2}{1+\alpha}}
}
for $x \ll 1/8$, where $\alpha \approx \log G^{2/3} \log \log G^{2/3}$ for $G=n/(Na) \gg 1$. This expression neglects critical effects relevant as $x \to 1/8^-$, where $\alpha
\to e$. 

Eq.\ref{p} predicts the observations of Fig.\ref{fig3}. First, since $G \propto n$ we obtain the prediction $p \propto n^{\beta(x)}$ as observed in Fig.\ref{fig3}a, with $\beta(x) = 2/(1+\alpha)$, rationalizing the observed continuously changing exponent. We predict $\beta(1/8) \approx 2/(1+e) = 0.538$ while $\beta(x) \sim
3/\log G \to 0$ at small~$x$ (large $G$). Second, the prefactor $a \sim \sqrt{8x}$ explains the small-$x$ scaling $p \propto \sqrt{x}$ observed in Fig.\ref{fig3}c. Third, the prefactor $\log(\alpha)$ explains the $N$ dependence of $p \propto N \log \log G$ at large $G$, observed in Fig.\ref{fig3}c.

\paragraph{Entropy, Energy, and Code length} From the dominant pattern we obtain an expression for $\ZZ$, sketched in End Matter, and extract its energetic ($E$) and entropic ($S_{\mathrm{Bol}}$) parts. The part of $S_{\mathrm{Bol}}$ that is extensive in $n$ in the large $n$ limit gives the entropy rate $H_d$ of (leftmost) derivations; with a deterministic surface layer, this is equal to the entropy rate of observed text. The corrections to strict extensivity encode the correlations in finite corpora. 
%Using \eqref{p} and its extension to other regimes, we finally derive expressions for the energy and entropy \cite{De-Giuli26}. We focus on the entropy rate $H_d$ of (leftmost) derivations; with a deterministic surface layer, this is equal to the entropy rate of observed text. 
For $H_d$ we predict, for $x_{\mathrm{lo}} < x < 1/8$, 
\eq{ \label{Hhi}
H_{\mathrm{hi}}(x) \approx \frac{1}{4} \log 8x + \frac{3k_*/2}{(k_*-a)^2} + \text{const} %\qquad 
}
where $k_* \approx \ffrac{3}{4} + \half a +
\half \sqrt{a^2 + a + \ffrac{9}{4}}$.
We predict a logarithmic behavior at small $x$ with prefactor $1/4$, verified in
\cref{fig4}a (dash-dotted and dotted lines), and a steepening as $x\to 1/8^-$
(dashed line), also captured by theory; for very small $x<x_{\mathrm{lo}}$ (not accessed in
numerics) we predict that~$H_d$ saturates. %a crossover to $H_{\mathrm{lo}}(x) \approx \frac{7}{8} C (8x)^{3/8} + \text{const}$. 
For the energy, subtracting a constant, we predict for the
hidden part $E_d$
\eq{
E_d-E_0/\ell = - \begin{cases} 1/(2\ed) & x \gtrsim 1/8 \\ 
\sqrt{w/2\ed} + \ldots& x < 1/8 \end{cases} ,
}
where we neglect subleading terms. This is verified in \cref{fig4}b and
decisively shows the phase transition at $x_c=1/8$, where the two branches intersect.

This completes our theoretical description of the RLM
summarized in the phase diagram of \cref{fig:phase-diagram}. In the following, we discuss two applications of the RLM: we demonstrate the emergence of Heaps
law~\cite{Heaps78}, and we analyse the dependence of entropy on context length
in LLMs~\cite{Scheibner25, cagnetta2026deriving}.

\paragraph{Application to natural language:} An empirical result, due to Heaps,
is that the number of unique {\it words} $n_*$ in a text of length $\ell$
increases as a power $n_* \sim \ell^b$ where $b \approx 1/2$
\cite{Heaps78,Petersen12}. This is recovered here since we predict $b \approx 1/2$ in the transition regime and the
critical regime, see \cite{De-Giuli26}. % SG: I removed the ref to \beta, since it is not defined anywhere else in the Letter. 
This suggests that natural languages exist near the condensation transition at $x_c=1/8$. Heaps' law is also coincident with Zipf's law
\cite{Zipf13,Ferrer-i-Cancho01,Corominas-Murtra10,Corral15}, see
\cite{De-Giuli26}. %\SG{Expand on the implictations on the ``temperature'' of RLM.}

\paragraph{Application to Large Language Models:} LLMs allow one to probe text probabilities as a function of context length $K$. With the semi-annealed theory, we can obtain the entropy rate as a function of corpus length~$n$, in the large $n$ regime (formally the scaling limit is taken at fixed $n/N$).
%by construction this is the largest possible correlation length of the text, according to the model. 
Ref.\cite{Scheibner25} measured the code length $\mathcal L = -\langle \log_2
P(x_0 | x_{-K},x_{-K+1},\ldots x_{-1}) \rangle$ as a function of $K$, which is
an approximation to the entropy rate $H$ computed on blocks of $K$ tokens, as
checked directly \cite{Scheibner25}; see also Ref. \cite{cagnetta2026deriving}. We reproduce the result for OLMo 2
\cite{OLMo24} in \cref{fig4}c (red curve). Assuming the data is in the frozen
phase, the semi-annealed theory predicts the leading behavior $\mathcal{L}
\approx \text{const} - \frac{1}{2} \log \alpha(G)$, and hence a very slow decay
$\sim -\log \log G$ towards the asymptotic limit. Here $G=K/Na$ and we fit
$Na\approx 2.2$ and the additive constant. A factor $1/n_{c/t}$, where
$n_{c/t}=4.79$ is the number of characters per token \cite{Scheibner25},
converts the RLM prediction to an entropy rate per character. As shown in
\cref{fig4}c (blue), this quantitatively matches the data for the entire range
over which the saddle exists,  $G \gtrsim 59$; at smaller $G$, the underlying
saddle of the semi-annealed theory disappears, and presumably a treatment of
finite sums is required. 

\paragraph{Conclusion:} In the scaling limit, the Random Language
Model exhibits a series of transitions where structure emerges, starting with pair-wise correlations and non-uniform marginals before strong structure emerges at a REM-like condensation transition at
$x_c=1/8$. The scaling limit is relevant linguistically: annotating hidden symbols with features makes their number explode combinatorially. Yet the slow approach to this limit explains why finite systems display
smooth crossovers. Our results settle the
debate about the existence of a phase transition in the
RLM~\cite{Nakaishi24,Toji26} and highlight that structure beyond CFG, while relevant
linguistically, is not necessary to create one.
%In the scaling limit, the Random Language Model has a REM-like freezing transition at $x_c=1/8$. Due to the glacial approach to the thermodynamic limit, any finite N behaves similarly, showing both babbling and frozen phases, yet connected by a smooth crossover. This settles the debate over the existence of a phase transition. 
A semi-annealed approximation reveals the role of rule multiplicity and explains hidden scaling laws in rule use over corpus length, rescaled temperature, and lexicon size. These predict the behavior of energy and entropy in the model. The RLM predicts both classical universal properties of natural language, such as Heaps' law, and quantitatively matches the dependence of code length on context length in LLMs.

Future work should connect these results to neural scaling laws \cite{Kaplan20}; for a data-driven theory, see \cite{cagnetta2026deriving}. Finally, the semi-annealed construction is not specific to language, but applies more broadly to systems in which macroscopic behavior is controlled by rare configurations selected from a disordered ensemble. Examples include glassy systems, optimization landscapes in machine learning, and evolutionary fitness landscapes, where dominant states arise from extreme-value statistics but must be described in terms of coarse-grained, permutation-invariant observables. 

\paragraph{Acknowledgments} EDG is grateful to Alessandro Treves, and SG is
grateful to Francesco Cagnetta, for stimulating conversations. AG acknowledges
funding from Next Generation EU, in the context of the Italian National Recovery
and Resilience Plan (PNRR), Missione 4, Componente 1, Investimento 4.1
``Estensione del numero di dottorati di ricerca e dottorati innovativi per la
Pubblica Amministrazione e il patrimonio culturale'' (CUP G93C23000620003). EDG
is supported by NSERC Discovery Grant RGPIN-2020-04762. SG acknowledges funding
from the European Research Council (ERC) for the project ``beyond2'' under the
European Union’s Horizon 2020 research and innovation programme, Grant agreement
ID 101166056, and from Next Generation EU, in the context of the National
Recovery and Resilience Plan, Investment PE1 – Project FAIR ``Future Artificial
Intelligence Research'' (CUP G53C22000440006).

\color{black}
%\cleardoublepage
%\twocolumn
%\balance
%\bibliographystyle{ieeetr}
%\addcontentsline{toc}{chapter}{Bibliography}
\bibliography{Language,Glasses}

\begin{thebibliography}{41}%
\makeatletter
\providecommand \@ifxundefined [1]{%
 \@ifx{#1\undefined}
}%
\providecommand \@ifnum [1]{%
 \ifnum #1\expandafter \@firstoftwo
 \else \expandafter \@secondoftwo
 \fi
}%
\providecommand \@ifx [1]{%
 \ifx #1\expandafter \@firstoftwo
 \else \expandafter \@secondoftwo
 \fi
}%
\providecommand \natexlab [1]{#1}%
\providecommand \enquote  [1]{``#1''}%
\providecommand \bibnamefont  [1]{#1}%
\providecommand \bibfnamefont [1]{#1}%
\providecommand \citenamefont [1]{#1}%
\providecommand \href@noop [0]{\@secondoftwo}%
\providecommand \href [0]{\begingroup \@sanitize@url \@href}%
\providecommand \@href[1]{\@@startlink{#1}\@@href}%
\providecommand \@@href[1]{\endgroup#1\@@endlink}%
\providecommand \@sanitize@url [0]{\catcode `\\12\catcode `\$12\catcode
  `\&12\catcode `\#12\catcode `\^12\catcode `\_12\catcode `\%12\relax}%
\providecommand \@@startlink[1]{}%
\providecommand \@@endlink[0]{}%
\providecommand \url  [0]{\begingroup\@sanitize@url \@url }%
\providecommand \@url [1]{\endgroup\@href {#1}{\urlprefix }}%
\providecommand \urlprefix  [0]{URL }%
\providecommand \Eprint [0]{\href }%
\providecommand \doibase [0]{http://dx.doi.org/}%
\providecommand \selectlanguage [0]{\@gobble}%
\providecommand \bibinfo  [0]{\@secondoftwo}%
\providecommand \bibfield  [0]{\@secondoftwo}%
\providecommand \translation [1]{[#1]}%
\providecommand \BibitemOpen [0]{}%
\providecommand \bibitemStop [0]{}%
\providecommand \bibitemNoStop [0]{.\EOS\space}%
\providecommand \EOS [0]{\spacefactor3000\relax}%
\providecommand \BibitemShut  [1]{\csname bibitem#1\endcsname}%
\let\auto@bib@innerbib\@empty
%</preamble>
\bibitem [{\citenamefont {De~Giuli}(2019{\natexlab{a}})}]{DeGiuli19}%
  \BibitemOpen
  \bibfield  {author} {\bibinfo {author} {\bibfnamefont {E.}~\bibnamefont
  {De~Giuli}},\ }\href@noop {} {\bibfield  {journal} {\bibinfo  {journal}
  {Phys. Rev. Lett.}\ }\textbf {\bibinfo {volume} {122}},\ \bibinfo {pages}
  {128301} (\bibinfo {year} {2019}{\natexlab{a}})}\BibitemShut {NoStop}%
\bibitem [{\citenamefont {Kaplan}\ \emph {et~al.}(2020)\citenamefont {Kaplan},
  \citenamefont {McCandlish}, \citenamefont {Henighan}, \citenamefont {Brown},
  \citenamefont {Chess}, \citenamefont {Child}, \citenamefont {Gray},
  \citenamefont {Radford}, \citenamefont {Wu},\ and\ \citenamefont
  {Amodei}}]{Kaplan20}%
  \BibitemOpen
  \bibfield  {author} {\bibinfo {author} {\bibfnamefont {J.}~\bibnamefont
  {Kaplan}}, \bibinfo {author} {\bibfnamefont {S.}~\bibnamefont {McCandlish}},
  \bibinfo {author} {\bibfnamefont {T.}~\bibnamefont {Henighan}}, \bibinfo
  {author} {\bibfnamefont {T.~B.}\ \bibnamefont {Brown}}, \bibinfo {author}
  {\bibfnamefont {B.}~\bibnamefont {Chess}}, \bibinfo {author} {\bibfnamefont
  {R.}~\bibnamefont {Child}}, \bibinfo {author} {\bibfnamefont
  {S.}~\bibnamefont {Gray}}, \bibinfo {author} {\bibfnamefont {A.}~\bibnamefont
  {Radford}}, \bibinfo {author} {\bibfnamefont {J.}~\bibnamefont {Wu}}, \ and\
  \bibinfo {author} {\bibfnamefont {D.}~\bibnamefont {Amodei}},\ }\href@noop {}
  {\bibfield  {journal} {\bibinfo  {journal} {arXiv preprint arXiv:2001.08361}\
  } (\bibinfo {year} {2020})}\BibitemShut {NoStop}%
\bibitem [{\citenamefont {De~Giuli}(2019{\natexlab{b}})}]{DeGiuli19a}%
  \BibitemOpen
  \bibfield  {author} {\bibinfo {author} {\bibfnamefont {E.}~\bibnamefont
  {De~Giuli}},\ }\href@noop {} {\bibfield  {journal} {\bibinfo  {journal}
  {Journal of Physics A: Mathematical and Theoretical}\ }\textbf {\bibinfo
  {volume} {52}},\ \bibinfo {pages} {504001} (\bibinfo {year}
  {2019}{\natexlab{b}})}\BibitemShut {NoStop}%
\bibitem [{\citenamefont {De~Giuli}(2022)}]{De-Giuli22}%
  \BibitemOpen
  \bibfield  {author} {\bibinfo {author} {\bibfnamefont {E.}~\bibnamefont
  {De~Giuli}},\ }\href@noop {} {\bibfield  {journal} {\bibinfo  {journal}
  {Journal of Physics A: Mathematical and Theoretical}\ }\textbf {\bibinfo
  {volume} {55}},\ \bibinfo {pages} {489501} (\bibinfo {year}
  {2022})}\BibitemShut {NoStop}%
\bibitem [{\citenamefont {Nakaishi}\ and\ \citenamefont
  {Hukushima}(2022)}]{Nakaishi22}%
  \BibitemOpen
  \bibfield  {author} {\bibinfo {author} {\bibfnamefont {K.}~\bibnamefont
  {Nakaishi}}\ and\ \bibinfo {author} {\bibfnamefont {K.}~\bibnamefont
  {Hukushima}},\ }\href@noop {} {\bibfield  {journal} {\bibinfo  {journal}
  {Physical Review Research}\ }\textbf {\bibinfo {volume} {4}},\ \bibinfo
  {pages} {023156} (\bibinfo {year} {2022})}\BibitemShut {NoStop}%
\bibitem [{\citenamefont {Lalegani}\ and\ \citenamefont
  {De~Giuli}(2024)}]{Lalegani24}%
  \BibitemOpen
  \bibfield  {author} {\bibinfo {author} {\bibfnamefont {F.}~\bibnamefont
  {Lalegani}}\ and\ \bibinfo {author} {\bibfnamefont {E.}~\bibnamefont
  {De~Giuli}},\ }\href@noop {} {\bibfield  {journal} {\bibinfo  {journal}
  {Physical Review E}\ }\textbf {\bibinfo {volume} {109}},\ \bibinfo {pages}
  {054313} (\bibinfo {year} {2024})}\BibitemShut {NoStop}%
\bibitem [{\citenamefont {Lin}\ and\ \citenamefont {Tegmark}(2017)}]{Lin17}%
  \BibitemOpen
  \bibfield  {author} {\bibinfo {author} {\bibfnamefont {H.~W.}\ \bibnamefont
  {Lin}}\ and\ \bibinfo {author} {\bibfnamefont {M.}~\bibnamefont {Tegmark}},\
  }\href@noop {} {\bibfield  {journal} {\bibinfo  {journal} {Entropy}\ }\textbf
  {\bibinfo {volume} {19}},\ \bibinfo {pages} {299} (\bibinfo {year}
  {2017})}\BibitemShut {NoStop}%
\bibitem [{\citenamefont {Cagnetta}\ \emph {et~al.}(2024)\citenamefont
  {Cagnetta}, \citenamefont {Petrini}, \citenamefont {Tomasini}, \citenamefont
  {Favero},\ and\ \citenamefont {Wyart}}]{Cagnetta24}%
  \BibitemOpen
  \bibfield  {author} {\bibinfo {author} {\bibfnamefont {F.}~\bibnamefont
  {Cagnetta}}, \bibinfo {author} {\bibfnamefont {L.}~\bibnamefont {Petrini}},
  \bibinfo {author} {\bibfnamefont {U.~M.}\ \bibnamefont {Tomasini}}, \bibinfo
  {author} {\bibfnamefont {A.}~\bibnamefont {Favero}}, \ and\ \bibinfo {author}
  {\bibfnamefont {M.}~\bibnamefont {Wyart}},\ }\href@noop {} {\bibfield
  {journal} {\bibinfo  {journal} {Physical Review X}\ }\textbf {\bibinfo
  {volume} {14}},\ \bibinfo {pages} {031001} (\bibinfo {year}
  {2024})}\BibitemShut {NoStop}%
\bibitem [{\citenamefont {Cagnetta}\ and\ \citenamefont
  {Wyart}(2024)}]{Cagnetta24a}%
  \BibitemOpen
  \bibfield  {author} {\bibinfo {author} {\bibfnamefont {F.}~\bibnamefont
  {Cagnetta}}\ and\ \bibinfo {author} {\bibfnamefont {M.}~\bibnamefont
  {Wyart}},\ }\href@noop {} {\bibfield  {journal} {\bibinfo  {journal}
  {Advances in Neural Information Processing Systems}\ }\textbf {\bibinfo
  {volume} {37}},\ \bibinfo {pages} {83119} (\bibinfo {year}
  {2024})}\BibitemShut {NoStop}%
\bibitem [{\citenamefont {Garnier-Brun}\ \emph {et~al.}(2024)\citenamefont
  {Garnier-Brun}, \citenamefont {M{\'e}zard}, \citenamefont {Moscato},\ and\
  \citenamefont {Saglietti}}]{Garnier-Brun24}%
  \BibitemOpen
  \bibfield  {author} {\bibinfo {author} {\bibfnamefont {J.}~\bibnamefont
  {Garnier-Brun}}, \bibinfo {author} {\bibfnamefont {M.}~\bibnamefont
  {M{\'e}zard}}, \bibinfo {author} {\bibfnamefont {E.}~\bibnamefont {Moscato}},
  \ and\ \bibinfo {author} {\bibfnamefont {L.}~\bibnamefont {Saglietti}},\
  }\href@noop {} {\bibfield  {journal} {\bibinfo  {journal} {arXiv preprint
  arXiv:2408.15138}\ } (\bibinfo {year} {2024})}\BibitemShut {NoStop}%
\bibitem [{\citenamefont {Cagnetta}\ \emph {et~al.}(2026)\citenamefont
  {Cagnetta}, \citenamefont {Ravent{\'o}s}, \citenamefont {Ganguli},\ and\
  \citenamefont {Wyart}}]{cagnetta2026deriving}%
  \BibitemOpen
  \bibfield  {author} {\bibinfo {author} {\bibfnamefont {F.}~\bibnamefont
  {Cagnetta}}, \bibinfo {author} {\bibfnamefont {A.}~\bibnamefont
  {Ravent{\'o}s}}, \bibinfo {author} {\bibfnamefont {S.}~\bibnamefont
  {Ganguli}}, \ and\ \bibinfo {author} {\bibfnamefont {M.}~\bibnamefont
  {Wyart}},\ }\href@noop {} {\bibfield  {journal} {\bibinfo  {journal} {arXiv
  preprint arXiv:2602.07488}\ } (\bibinfo {year} {2026})}\BibitemShut {NoStop}%
\bibitem [{\citenamefont {Chomsky}(2002)}]{Chomsky02}%
  \BibitemOpen
  \bibfield  {author} {\bibinfo {author} {\bibfnamefont {N.}~\bibnamefont
  {Chomsky}},\ }\href@noop {} {\emph {\bibinfo {title} {Syntactic
  structures}}}\ (\bibinfo  {publisher} {Walter de Gruyter},\ \bibinfo
  {address} {Berlin},\ \bibinfo {year} {2002})\BibitemShut {NoStop}%
\bibitem [{\citenamefont {Chomsky}(2014)}]{Chomsky14}%
  \BibitemOpen
  \bibfield  {author} {\bibinfo {author} {\bibfnamefont {N.}~\bibnamefont
  {Chomsky}},\ }\href@noop {} {\emph {\bibinfo {title} {Aspects of the Theory
  of Syntax}}},\ Vol.~\bibinfo {volume} {11}\ (\bibinfo  {publisher} {MIT
  press},\ \bibinfo {address} {Cambridge},\ \bibinfo {year} {2014})\BibitemShut
  {NoStop}%
\bibitem [{\citenamefont {Hopcroft}\ \emph {et~al.}(2007)\citenamefont
  {Hopcroft}, \citenamefont {Motwani},\ and\ \citenamefont
  {Ullman}}]{Hopcroft07}%
  \BibitemOpen
  \bibfield  {author} {\bibinfo {author} {\bibfnamefont {J.~E.}\ \bibnamefont
  {Hopcroft}}, \bibinfo {author} {\bibfnamefont {R.}~\bibnamefont {Motwani}}, \
  and\ \bibinfo {author} {\bibfnamefont {J.~D.}\ \bibnamefont {Ullman}},\
  }\href@noop {} {\emph {\bibinfo {title} {Introduction to automata theory,
  languages, and computation}}},\ \bibinfo {edition} {3rd}\ ed.\ (\bibinfo
  {publisher} {Pearson},\ \bibinfo {address} {Boston, Ma},\ \bibinfo {year}
  {2007})\BibitemShut {NoStop}%
\bibitem [{\citenamefont {Derrida}(1980)}]{Derrida80}%
  \BibitemOpen
  \bibfield  {author} {\bibinfo {author} {\bibfnamefont {B.}~\bibnamefont
  {Derrida}},\ }\href@noop {} {\bibfield  {journal} {\bibinfo  {journal}
  {Physical Review Letters}\ }\textbf {\bibinfo {volume} {45}},\ \bibinfo
  {pages} {79} (\bibinfo {year} {1980})}\BibitemShut {NoStop}%
\bibitem [{\citenamefont {Derrida}(1981)}]{Derrida81}%
  \BibitemOpen
  \bibfield  {author} {\bibinfo {author} {\bibfnamefont {B.}~\bibnamefont
  {Derrida}},\ }\href {\doibase 10.1103/PhysRevB.24.2613} {\bibfield  {journal}
  {\bibinfo  {journal} {Phys. Rev. B}\ }\textbf {\bibinfo {volume} {24}},\
  \bibinfo {pages} {2613} (\bibinfo {year} {1981})}\BibitemShut {NoStop}%
\bibitem [{\citenamefont {Heaps}(1978)}]{Heaps78}%
  \BibitemOpen
  \bibfield  {author} {\bibinfo {author} {\bibfnamefont {H.~S.}\ \bibnamefont
  {Heaps}},\ }\href@noop {} {\emph {\bibinfo {title} {Information retrieval:
  Computational and theoretical aspects}}}\ (\bibinfo  {publisher} {Academic
  Press, Inc.},\ \bibinfo {year} {1978})\BibitemShut {NoStop}%
\bibitem [{\citenamefont {Scheibner}\ \emph {et~al.}(2025)\citenamefont
  {Scheibner}, \citenamefont {Smith},\ and\ \citenamefont
  {Bialek}}]{Scheibner25}%
  \BibitemOpen
  \bibfield  {author} {\bibinfo {author} {\bibfnamefont {C.}~\bibnamefont
  {Scheibner}}, \bibinfo {author} {\bibfnamefont {L.~M.}\ \bibnamefont
  {Smith}}, \ and\ \bibinfo {author} {\bibfnamefont {W.}~\bibnamefont
  {Bialek}},\ }\href@noop {} {\bibfield  {journal} {\bibinfo  {journal} {arXiv
  preprint arXiv:2512.24969}\ } (\bibinfo {year} {2025})}\BibitemShut {NoStop}%
\bibitem [{\citenamefont {De~Giuli}(2026)}]{De-Giuli26}%
  \BibitemOpen
  \bibfield  {author} {\bibinfo {author} {\bibfnamefont {E.}~\bibnamefont
  {De~Giuli}},\ }\href@noop {} {\enquote {\bibinfo {title} {Scaling limit of
  the {R}andom {L}anguage {M}odel},}\ } (\bibinfo {year} {2026})\BibitemShut
  {NoStop}%
\bibitem [{\citenamefont {M\'ezard}\ and\ \citenamefont
  {Montanari}(2009)}]{Mezard09}%
  \BibitemOpen
  \bibfield  {author} {\bibinfo {author} {\bibfnamefont {M.}~\bibnamefont
  {M\'ezard}}\ and\ \bibinfo {author} {\bibfnamefont {A.}~\bibnamefont
  {Montanari}},\ }\href@noop {} {\emph {\bibinfo {title} {Information, physics,
  and computation}}}\ (\bibinfo  {publisher} {Oxford University Press},\
  \bibinfo {year} {2009})\BibitemShut {NoStop}%
\bibitem [{\citenamefont {Mézard}\ \emph {et~al.}(2002)\citenamefont
  {Mézard}, \citenamefont {Parisi},\ and\ \citenamefont
  {Zecchina}}]{mezard2002analytic}%
  \BibitemOpen
  \bibfield  {author} {\bibinfo {author} {\bibfnamefont {M.}~\bibnamefont
  {Mézard}}, \bibinfo {author} {\bibfnamefont {G.}~\bibnamefont {Parisi}}, \
  and\ \bibinfo {author} {\bibfnamefont {R.}~\bibnamefont {Zecchina}},\ }\href
  {\doibase 10.1126/science.1073287} {\bibfield  {journal} {\bibinfo  {journal}
  {Science}\ }\textbf {\bibinfo {volume} {297}},\ \bibinfo {pages} {812}
  (\bibinfo {year} {2002})}\BibitemShut {NoStop}%
\bibitem [{\citenamefont {Braunstein}\ \emph {et~al.}(2005)\citenamefont
  {Braunstein}, \citenamefont {M{\'e}zard},\ and\ \citenamefont
  {Zecchina}}]{braunstein2005survey}%
  \BibitemOpen
  \bibfield  {author} {\bibinfo {author} {\bibfnamefont {A.}~\bibnamefont
  {Braunstein}}, \bibinfo {author} {\bibfnamefont {M.}~\bibnamefont
  {M{\'e}zard}}, \ and\ \bibinfo {author} {\bibfnamefont {R.}~\bibnamefont
  {Zecchina}},\ }\href@noop {} {\bibfield  {journal} {\bibinfo  {journal}
  {Random Structures \& Algorithms}\ }\textbf {\bibinfo {volume} {27}},\
  \bibinfo {pages} {201} (\bibinfo {year} {2005})}\BibitemShut {NoStop}%
\bibitem [{\citenamefont {Braunstein}\ and\ \citenamefont
  {Zecchina}(2006)}]{braunstein2006learning}%
  \BibitemOpen
  \bibfield  {author} {\bibinfo {author} {\bibfnamefont {A.}~\bibnamefont
  {Braunstein}}\ and\ \bibinfo {author} {\bibfnamefont {R.}~\bibnamefont
  {Zecchina}},\ }\href@noop {} {\bibfield  {journal} {\bibinfo  {journal}
  {Physical review letters}\ }\textbf {\bibinfo {volume} {96}},\ \bibinfo
  {pages} {030201} (\bibinfo {year} {2006})}\BibitemShut {NoStop}%
\bibitem [{\citenamefont {Zdeborov{\'a}}\ and\ \citenamefont
  {Krzkala}(2007)}]{zdeborova2007phase}%
  \BibitemOpen
  \bibfield  {author} {\bibinfo {author} {\bibfnamefont {L.}~\bibnamefont
  {Zdeborov{\'a}}}\ and\ \bibinfo {author} {\bibfnamefont {F.}~\bibnamefont
  {Krzkala}},\ }\href@noop {} {\bibfield  {journal} {\bibinfo  {journal}
  {Physical Review E—Statistical, Nonlinear, and Soft Matter Physics}\
  }\textbf {\bibinfo {volume} {76}},\ \bibinfo {pages} {031131} (\bibinfo
  {year} {2007})}\BibitemShut {NoStop}%
\bibitem [{\citenamefont {Decelle}\ \emph {et~al.}(2011)\citenamefont
  {Decelle}, \citenamefont {Krzakala}, \citenamefont {Moore},\ and\
  \citenamefont {Zdeborov{\'a}}}]{decelle2011inference}%
  \BibitemOpen
  \bibfield  {author} {\bibinfo {author} {\bibfnamefont {A.}~\bibnamefont
  {Decelle}}, \bibinfo {author} {\bibfnamefont {F.}~\bibnamefont {Krzakala}},
  \bibinfo {author} {\bibfnamefont {C.}~\bibnamefont {Moore}}, \ and\ \bibinfo
  {author} {\bibfnamefont {L.}~\bibnamefont {Zdeborov{\'a}}},\ }\href@noop {}
  {\bibfield  {journal} {\bibinfo  {journal} {Physical Review Letters}\
  }\textbf {\bibinfo {volume} {107}},\ \bibinfo {pages} {065701} (\bibinfo
  {year} {2011})}\BibitemShut {NoStop}%
\bibitem [{\citenamefont {Krzakala}\ \emph {et~al.}(2012)\citenamefont
  {Krzakala}, \citenamefont {M{\'e}zard}, \citenamefont {Sausset},
  \citenamefont {Sun},\ and\ \citenamefont
  {Zdeborov{\'a}}}]{krzakala2012statistical}%
  \BibitemOpen
  \bibfield  {author} {\bibinfo {author} {\bibfnamefont {F.}~\bibnamefont
  {Krzakala}}, \bibinfo {author} {\bibfnamefont {M.}~\bibnamefont
  {M{\'e}zard}}, \bibinfo {author} {\bibfnamefont {F.}~\bibnamefont {Sausset}},
  \bibinfo {author} {\bibfnamefont {Y.}~\bibnamefont {Sun}}, \ and\ \bibinfo
  {author} {\bibfnamefont {L.}~\bibnamefont {Zdeborov{\'a}}},\ }\href@noop {}
  {\bibfield  {journal} {\bibinfo  {journal} {Physical Review X}\ }\textbf
  {\bibinfo {volume} {2}},\ \bibinfo {pages} {021005} (\bibinfo {year}
  {2012})}\BibitemShut {NoStop}%
\bibitem [{\citenamefont {Zdeborov{\'a}}\ and\ \citenamefont
  {Krzakala}(2016)}]{zdeborova2016statistical}%
  \BibitemOpen
  \bibfield  {author} {\bibinfo {author} {\bibfnamefont {L.}~\bibnamefont
  {Zdeborov{\'a}}}\ and\ \bibinfo {author} {\bibfnamefont {F.}~\bibnamefont
  {Krzakala}},\ }\href@noop {} {\bibfield  {journal} {\bibinfo  {journal}
  {Advances in Physics}\ }\textbf {\bibinfo {volume} {65}},\ \bibinfo {pages}
  {453} (\bibinfo {year} {2016})}\BibitemShut {NoStop}%
\bibitem [{\citenamefont {Ben~Arous}\ \emph {et~al.}(2005)\citenamefont
  {Ben~Arous}, \citenamefont {Bogachev},\ and\ \citenamefont
  {Molchanov}}]{Ben-Arous05}%
  \BibitemOpen
  \bibfield  {author} {\bibinfo {author} {\bibfnamefont {G.}~\bibnamefont
  {Ben~Arous}}, \bibinfo {author} {\bibfnamefont {L.~V.}\ \bibnamefont
  {Bogachev}}, \ and\ \bibinfo {author} {\bibfnamefont {S.~A.}\ \bibnamefont
  {Molchanov}},\ }\href@noop {} {\bibfield  {journal} {\bibinfo  {journal}
  {Probability theory and related fields}\ }\textbf {\bibinfo {volume} {132}},\
  \bibinfo {pages} {579} (\bibinfo {year} {2005})}\BibitemShut {NoStop}%
\bibitem [{\citenamefont {Kosterlitz}\ and\ \citenamefont
  {Thouless}(1973)}]{Kosterlitz73}%
  \BibitemOpen
  \bibfield  {author} {\bibinfo {author} {\bibfnamefont {J.~M.}\ \bibnamefont
  {Kosterlitz}}\ and\ \bibinfo {author} {\bibfnamefont {D.~J.}\ \bibnamefont
  {Thouless}},\ }\href@noop {} {\bibfield  {journal} {\bibinfo  {journal}
  {Journal of Physics C: Solid State Physics}\ }\textbf {\bibinfo {volume}
  {6}},\ \bibinfo {pages} {1181} (\bibinfo {year} {1973})}\BibitemShut
  {NoStop}%
\bibitem [{\citenamefont {Kosterlitz}(1974)}]{Kosterlitz74}%
  \BibitemOpen
  \bibfield  {author} {\bibinfo {author} {\bibfnamefont {J.}~\bibnamefont
  {Kosterlitz}},\ }\href@noop {} {\bibfield  {journal} {\bibinfo  {journal}
  {Journal of Physics C: Solid State Physics}\ }\textbf {\bibinfo {volume}
  {7}},\ \bibinfo {pages} {1046} (\bibinfo {year} {1974})}\BibitemShut
  {NoStop}%
\bibitem [{\citenamefont {Petersen}\ \emph {et~al.}(2012)\citenamefont
  {Petersen}, \citenamefont {Tenenbaum}, \citenamefont {Havlin}, \citenamefont
  {Stanley},\ and\ \citenamefont {Perc}}]{Petersen12}%
  \BibitemOpen
  \bibfield  {author} {\bibinfo {author} {\bibfnamefont {A.~M.}\ \bibnamefont
  {Petersen}}, \bibinfo {author} {\bibfnamefont {J.~N.}\ \bibnamefont
  {Tenenbaum}}, \bibinfo {author} {\bibfnamefont {S.}~\bibnamefont {Havlin}},
  \bibinfo {author} {\bibfnamefont {H.~E.}\ \bibnamefont {Stanley}}, \ and\
  \bibinfo {author} {\bibfnamefont {M.}~\bibnamefont {Perc}},\ }\href@noop {}
  {\bibfield  {journal} {\bibinfo  {journal} {Scientific reports}\ }\textbf
  {\bibinfo {volume} {2}},\ \bibinfo {pages} {943} (\bibinfo {year}
  {2012})}\BibitemShut {NoStop}%
\bibitem [{\citenamefont {Zipf}(2013)}]{Zipf13}%
  \BibitemOpen
  \bibfield  {author} {\bibinfo {author} {\bibfnamefont {G.~K.}\ \bibnamefont
  {Zipf}},\ }\href@noop {} {\emph {\bibinfo {title} {The psycho-biology of
  language: An introduction to dynamic philology}}}\ (\bibinfo  {publisher}
  {Routledge},\ \bibinfo {address} {Milton Park},\ \bibinfo {year}
  {2013})\BibitemShut {NoStop}%
\bibitem [{\citenamefont {Ferrer~i Cancho}\ and\ \citenamefont
  {Sol{\'e}}(2001)}]{Ferrer-i-Cancho01}%
  \BibitemOpen
  \bibfield  {author} {\bibinfo {author} {\bibfnamefont {R.}~\bibnamefont
  {Ferrer~i Cancho}}\ and\ \bibinfo {author} {\bibfnamefont {R.~V.}\
  \bibnamefont {Sol{\'e}}},\ }\href@noop {} {\bibfield  {journal} {\bibinfo
  {journal} {Journal of Quantitative Linguistics}\ }\textbf {\bibinfo {volume}
  {8}},\ \bibinfo {pages} {165} (\bibinfo {year} {2001})}\BibitemShut {NoStop}%
\bibitem [{\citenamefont {Corominas-Murtra}\ and\ \citenamefont
  {Sol{\'e}}(2010)}]{Corominas-Murtra10}%
  \BibitemOpen
  \bibfield  {author} {\bibinfo {author} {\bibfnamefont {B.}~\bibnamefont
  {Corominas-Murtra}}\ and\ \bibinfo {author} {\bibfnamefont {R.~V.}\
  \bibnamefont {Sol{\'e}}},\ }\href@noop {} {\bibfield  {journal} {\bibinfo
  {journal} {Physical Review E}\ }\textbf {\bibinfo {volume} {82}},\ \bibinfo
  {pages} {011102} (\bibinfo {year} {2010})}\BibitemShut {NoStop}%
\bibitem [{\citenamefont {Corral}\ \emph {et~al.}(2015)\citenamefont {Corral},
  \citenamefont {Boleda},\ and\ \citenamefont {Ferrer-i Cancho}}]{Corral15}%
  \BibitemOpen
  \bibfield  {author} {\bibinfo {author} {\bibfnamefont {A.}~\bibnamefont
  {Corral}}, \bibinfo {author} {\bibfnamefont {G.}~\bibnamefont {Boleda}}, \
  and\ \bibinfo {author} {\bibfnamefont {R.}~\bibnamefont {Ferrer-i Cancho}},\
  }\href@noop {} {\bibfield  {journal} {\bibinfo  {journal} {PloS one}\
  }\textbf {\bibinfo {volume} {10}},\ \bibinfo {pages} {e0129031} (\bibinfo
  {year} {2015})}\BibitemShut {NoStop}%
\bibitem [{\citenamefont {OLMo}\ \emph {et~al.}(2024)\citenamefont {OLMo},
  \citenamefont {Walsh}, \citenamefont {Soldaini}, \citenamefont {Groeneveld},
  \citenamefont {Lo}, \citenamefont {Arora}, \citenamefont {Bhagia},
  \citenamefont {Gu}, \citenamefont {Huang},\ and\ \citenamefont
  {Jordan}}]{OLMo24}%
  \BibitemOpen
  \bibfield  {author} {\bibinfo {author} {\bibfnamefont {T.}~\bibnamefont
  {OLMo}}, \bibinfo {author} {\bibfnamefont {P.}~\bibnamefont {Walsh}},
  \bibinfo {author} {\bibfnamefont {L.}~\bibnamefont {Soldaini}}, \bibinfo
  {author} {\bibfnamefont {D.}~\bibnamefont {Groeneveld}}, \bibinfo {author}
  {\bibfnamefont {K.}~\bibnamefont {Lo}}, \bibinfo {author} {\bibfnamefont
  {S.}~\bibnamefont {Arora}}, \bibinfo {author} {\bibfnamefont
  {A.}~\bibnamefont {Bhagia}}, \bibinfo {author} {\bibfnamefont
  {Y.}~\bibnamefont {Gu}}, \bibinfo {author} {\bibfnamefont {S.}~\bibnamefont
  {Huang}}, \ and\ \bibinfo {author} {\bibfnamefont {M.}~\bibnamefont
  {Jordan}},\ }\href@noop {} {\bibfield  {journal} {\bibinfo  {journal} {arXiv
  preprint arXiv:2501.00656}\ } (\bibinfo {year} {2024})}\BibitemShut {NoStop}%
\bibitem [{\citenamefont {Nakaishi}\ and\ \citenamefont
  {Hukushima}(2024)}]{Nakaishi24}%
  \BibitemOpen
  \bibfield  {author} {\bibinfo {author} {\bibfnamefont {K.}~\bibnamefont
  {Nakaishi}}\ and\ \bibinfo {author} {\bibfnamefont {K.}~\bibnamefont
  {Hukushima}},\ }\href@noop {} {\bibfield  {journal} {\bibinfo  {journal}
  {Physical Review Research}\ }\textbf {\bibinfo {volume} {6}},\ \bibinfo
  {pages} {033216} (\bibinfo {year} {2024})}\BibitemShut {NoStop}%
\bibitem [{\citenamefont {Toji}\ \emph {et~al.}(2026)\citenamefont {Toji},
  \citenamefont {Takahashi}, \citenamefont {Roychowdhury},\ and\ \citenamefont
  {Miyahara}}]{Toji26}%
  \BibitemOpen
  \bibfield  {author} {\bibinfo {author} {\bibfnamefont {Y.}~\bibnamefont
  {Toji}}, \bibinfo {author} {\bibfnamefont {J.}~\bibnamefont {Takahashi}},
  \bibinfo {author} {\bibfnamefont {V.}~\bibnamefont {Roychowdhury}}, \ and\
  \bibinfo {author} {\bibfnamefont {H.}~\bibnamefont {Miyahara}},\ }\href@noop
  {} {\bibfield  {journal} {\bibinfo  {journal} {Physical Review E}\ }\textbf
  {\bibinfo {volume} {113}},\ \bibinfo {pages} {015305} (\bibinfo {year}
  {2026})}\BibitemShut {NoStop}%
\bibitem [{\citenamefont {Cover}\ and\ \citenamefont {Thomas}(1999)}]{Cover99}%
  \BibitemOpen
  \bibfield  {author} {\bibinfo {author} {\bibfnamefont {T.~M.}\ \bibnamefont
  {Cover}}\ and\ \bibinfo {author} {\bibfnamefont {J.~A.}\ \bibnamefont
  {Thomas}},\ }\href@noop {} {\emph {\bibinfo {title} {Elements of information
  theory}}}\ (\bibinfo  {publisher} {John Wiley \& Sons},\ \bibinfo {year}
  {1999})\BibitemShut {NoStop}%
\bibitem [{\citenamefont {Grassberger}(2003)}]{Grassberger03}%
  \BibitemOpen
  \bibfield  {author} {\bibinfo {author} {\bibfnamefont {P.}~\bibnamefont
  {Grassberger}},\ }\href@noop {} {\bibfield  {journal} {\bibinfo  {journal}
  {arXiv preprint physics/0307138}\ } (\bibinfo {year} {2003})}\BibitemShut
  {NoStop}%
\bibitem [{\citenamefont {Sch{\"u}rmann}\ and\ \citenamefont
  {Grassberger}(1996)}]{Schurmann96}%
  \BibitemOpen
  \bibfield  {author} {\bibinfo {author} {\bibfnamefont {T.}~\bibnamefont
  {Sch{\"u}rmann}}\ and\ \bibinfo {author} {\bibfnamefont {P.}~\bibnamefont
  {Grassberger}},\ }\href@noop {} {\bibfield  {journal} {\bibinfo  {journal}
  {Chaos: An Interdisciplinary Journal of Nonlinear Science}\ }\textbf
  {\bibinfo {volume} {6}},\ \bibinfo {pages} {414} (\bibinfo {year}
  {1996})}\BibitemShut {NoStop}%
\end{thebibliography}%

\section{End Matter}

{\it Combinatorics: } Here we outline the combinatorics of patterns; for a complete treatment, including surface terms, see \cite{De-Giuli26}. First choose the symbols used by the $p$ branching rules; this gives $\binom{N^3}{p} \sim N^{3p}/p!$ if $N^3 \gg p$. Then distribute these according to the rarity distribution; this gives a multinomial coefficient $\binom{p}{p_1 \; p_2 \; \cdots }$. Having fixed the identities of the rules, we can create different derivations by placing them in different configurations. This can be estimated due to two simplifying factors of the model: first, the tree structure allows configurations to be built sequentially, starting from the root. Second, since rules are iid, if for example we have a rule $a \to bc$, then the likelihood of having a rule whose head is $b$ is independent of the value $(a)$ of the head. Suppose we have a pile of $\ell_c$ rules, of which $p$ of them are unique. We assume that $\ell_c/p$ is large, so that a typical rule appears frequently. Then, starting from the root, we will have typically $p/N$ rules with the start symbol as head. Continuing to its children, we will have $p/N$ choices for the next branch, and so on. This yields a simple estimate $(p/N)^{\ell_c}$ for the number of unique derivations. %It is confirmed by an calculation based on Eqs.XX in the annealed approximation. 
This implies the pile will yield a unique derivation if $p \approx N$.

However, this estimate does not account for the depletion of
rules as we build derivations. We account for this by dividing up the full $n$
rules (counted with multiplicity) into piles of $\ell_c$ rules, each of which
has $N$ unique rules. Each pile will then typically have a unique derivation.
Since the average repetition rate of a rule is $n/p$ we take $\ell_c = n N/p$,
and hence $m_c = p/N$ piles. This defines a self-consistent decomposition into
$m_c$ correlation volumes of size $\ell_c$. This requires $p/N\geq 1$, since
$m_c$ must be integer. The case $p<N$ is treated in \cite{De-Giuli26}.

For $p>N$, counting derivations thus reduces to counting the number of ways to partition the initial $n$ rules (with given multiplicity spectrum) into $m_c$ correlation volumes. Label the volumes 1 through $m_c$ and consider each rule type $r$, i.e. triple $r= (a,b,c)$ corresponding to $a \to bc$. If this rule occurs $\pi$ times in the complete pile of $n$ rules, then there are $\binom{\pi + m_c - 1}{m_c -1}$ ways to put these $\pi$ rules into the correlation volumes. Repeating this for all rules and recalling the definitions of $\{ p_k \}$ we find \footnote{This does not enforce that each pile has exactly $\ell_c$ rules, counted with multiplicity, although in the large $m_c$ limit this is the most likely outcome. The corrections are computable; see \cite{De-Giuli26}. }
$\prod_{k \geq 1} \binom{k + m_c - 1}{m_c -1}^{p_k} $. As a result, counting patterns gives the exponent $Q=3p$ and a factor
\eq{
\Omega = \frac{1}{p!} \binom{p}{p_1 \; p_2 \; \cdots } \prod_{k \geq 1} \binom{k + m_c - 1}{m_c -1}^{p_k} .
}

To complete the evaluation of $\mathcal{P}$, we need $\langle e^{k X} \rangle_{X<X_*}$. Using extreme-value statistics, a sum over $N^2$ variables sampled from $P(X)$ will typically not see a value larger than $X^*$ where
\eq{
\frac{1}{N^2}  = \int_{X_*}^\infty  dX P(X)  = \half \text{erfc}(\sqrt{\ed} X_*) .
}
Define an average restricted to the interval $X<X_*$:
\eqs{
%\langle f(X) \rangle_{X<X_*} = \frac{\int_{-\infty}^{X_*} dX P(X) f(X)}{\int_{-\infty}^{X_*} dX P(X)}
\langle f(X) \rangle_{X<X_*} = \left.\int_{-\infty}^{X_*} dX P(X) f(X) \right./  \int_{-\infty}^{X_*} dX P(X)
}
In the frozen phase $\langle e^{k X} \rangle_{X<X_*}~\sim~N^{-2 + k
\sqrt{w/2x\log N}} a/(k-a)$ with $a= \sqrt{2\ed w} \sim \sqrt{8x}$ and $w =
W_0(N^4/2\pi)$, where $W_0$ is Lambert's $W_0$. Up to subexponential factors
(retained in the full theory) each finite pattern has an exponential weight
$N^{p + n \sqrt{2/x}}$ that depends on $x$. Much like in the XY model
\cite{Kosterlitz73,Kosterlitz74}, the relevant excitations can depend on
temperature. It implies there is a preferred rule spectrum at any given $x$. We
find $p_k = v \binom{k + m_c - 1}{m_c -1} N^3 e^{- k \tilde n} \langle e^{k X}
\rangle_{X<X_*}$ with $v = e^{\sum_{k'} \frac{p_{k'}}{N} \log
\left(1+\frac{k'}{m_c-1}\right)}$ and where $\tilde n$ is a Lagrange multiplier
fixing $n=\sum_k k p_k$, which needs to be solved implicitly. Note also that $p=\sum_k p_k$. The result agrees
with direct measurements \cite{De-Giuli26} and predicts a smooth crossover from
nearly unique rule use at $x>1/8$ to a nontrivial spectrum at small $x$.

{\it Partition function: } Combining Eqs.\ref{Z},\ref{P},\ref{p} we obtain $\ZZ \approx \ZZ_0 e^{n X_* + p}/(v^p z^n)$ where $z \approx (3N/2p) \alpha/ \log \alpha$. Using the scaling symmetry of the model \cite{DeGiuli19a} this can be separated into energetic ($E$) and entropic ($S_{\mathrm{Bol}}$) parts: up to constant parts (see \cite{De-Giuli26}) we make the replacement $\ed \to \ed/\beta^2$ and then extract
\eqs{
\langle E \rangle & = -\p \langle \log \ZZ \rangle/\p \beta|_{\beta=1} \\
\langle S_{\mathrm{Bol}} \rangle & = \langle \log \ZZ \rangle + \langle E \rangle .
}
$S_{\mathrm{Bol}}$ is equal to the Shannon entropy of derivations in nats; for a deterministic surface layer, this is equal to the Shannon entropy of the hidden symbols only. Since there are $2n+m$ hidden symbols in a forest with $n$ branches, the hidden entropy rate is defined by $H_d = S_{\mathrm{Bol}}/(2n+m)$. Similarly, for a deterministic surface layer, $E_d = E/n$. In the frozen phase, $p \ll n$ and the leading terms come from $\log \ZZ = n X_* - n \log z + \ldots$. The $X_*$ term contributes to the energy while the $\log (\alpha/p)$ terms in $\log z$ contribute to the entropy. 

{\it Entropy rate: } The entropy rate $H_d$ is obtained theoretically from the RLM by the Boltzmann entropy, as described above. Numerically, however, it is obtained directly from the forest of derivations. Label the interior sites  $1,2,\ldots$ by a standard leftmost ordering \cite{Hopcroft07} and define (see section 4.2 of \cite{Cover99})
\begin{equation}
    \label{eq:Hd}
    H_d(\ell; m) = -\frac{1}{2n+m} \left\langle \log \PP(\sigma_1, \sigma_2, \ldots, \sigma_{2n+m} )\right\rangle .
\end{equation}
We use the technique of \cite{Grassberger03} to obtain unbiased estimates of $H_d$. Those reported in Fig. 4a use $\approx 120$ distinct grammars at each parameter value, and $\ell \approx 10^4$ for each grammar. Entropies are computed as differential entropies $\delta H_d = 2H_d(2)-H_d(1)$ (see \cite{Schurmann96}) over blocks of 2 adjacent symbols. 

The entropy rate of the observable sequences alone is
\eq{
H_s(\ell) = -\frac{1}{\ell} \left\langle \log \PP(o_1, o_2, \ldots, o_{\ell} )\right\rangle .
}
When the surface layer is deterministic and the grammar is unambiguous, $H_s=H_d$. As is well known, $H_s(\ell)$ converges, as $\ell \to \infty$, to the conditional entropy rate \cite{Schurmann96,Cover99}
\eq{
H_s'(\ell) = -\frac{1}{\ell} \left\langle \log \PP(o_\ell | o_1, o_2, \ldots, o_{\ell-1} )\right\rangle ,
%H_d' = -\frac{1}{2n} \left\langle \log \PP(\sigma_{2n} | \sigma_1, \sigma_2, \ldots, \sigma_{2n-1} )\right\rangle ,
}
whence its relation to the code length, which however does not require that the considered distribution is equal to the distribution from which the sequences are drawn.

\end{document}